\documentclass[onecolumn,floatfix,prl]{revtex4-1}

\usepackage{graphicx}
\usepackage{dcolumn}
\usepackage{bm}
\usepackage{amsmath,amssymb,mathtools,color}
\usepackage{color}

\def\b0{\mbox{\bf 0}}
\def\be{\mbox{{\bf e}}}

\def\bA{\mbox{{\bf A}}}
\def\bF{\mbox{{\bf F}}}
\def\bH{\mbox{{\bf H}}}

\def\bI{\mbox{\bf I}}

\def\bL{\mbox{{\bf L}}}

\begin{document}
\begin{abstract}
In a generalized framework, where multi-state and inter-state linkages are allowed, we derive a sufficient condition for the stability of synchronization in a network of chaotic attractors. This condition explicitly relates the network structure and the local and coupling dynamics to synchronization stability. For large Erd\"{o}s-R\'{e}nyi networks, the obtained condition is translated into a lower bound on the probability of stability of synchrony. Our results show that the probability of stability quickly increases as the randomness crosses a threshold which for large networks is inversely proportional to the network size.
\end{abstract}
\title{Synchronization Probability in Large Random Networks}
\author{Saeed~Manaffam and~Alireza~Seyedi\\\small Department of Electrical Engineering and Computer Science\\University of Central Florida}%

\date{\today}

\maketitle

\section{introduction}
The synchronization of complex dynamical systems is one of the most intriguing problems and has roots in physics, biology, and engineering \cite{Arenas08}. The problem of synchronization in a network of interacting nodes was first  brought to attention by Wiener \cite{Wiener48} and later pursued by Winfree, who modeled biological oscillators as phase oscillators, neglecting the amplitude \cite{Ariaratnam01}\cite{Winfree67}.

Following Winfree's pioneering work, there has been considerable effort to study the structural and dynamical effects of the network and its nodes on the state of synchrony. One important problem encountered in these studies is the possibility of separating the global (network) and local (node) characteristics to allow a general assessment on stability of the synchronous state. In the seminal work by Pecora and Carroll \cite{Pecora98}, this issue was addressed by introducing the concept of master stability equation. This equation provides a condition on stability of the network, known as master stability condition (MSC). The MSC also leads to useful bounds on structural properties of the network such as eigenvalues of the Laplacian matrix of the network \cite{Newman10}\cite{Pecora98}. Following this important result, most of the focus has been directed toward interpreting the bounds on eigenvalues of the Laplacian matrix of the network implied by the MSC as bounds on the degrees of nodes in the network. In \cite{Nishikawa03} and \cite{Wang02}, the effect of heterogeneity and \emph{smallness} of the network on synchronization stability have been investigated and it has been shown that if the network is more homogeneous the stability of synchrony is easier to achieve. Though most efforts have focused on the unweighted and undirected graphs, \cite{Motter07} and \cite{Zhou06} have investigated the master stability equation in directed and weighted structures, respectively.

Since the synchronization manifold can be considered as a fixed point of a reduced (induced) space, the vast majority of the existing literature on the synchronization in the networks considers the largest traversal Lyapunov exponent as a measure of stability of the synchronization. Of course, as noted in \cite{Pecora98}, the negativity of Lyapunov exponents is not a necessary nor a sufficient condition on the stability of manifold itself and it does not stop the manifold from bubbling and bursting \cite{Daleckii74}\cite{Leonov07}.

In this paper, we use an alternative master stability equation derived from Lyapunov direct method to obtain a condition on the stability of the whole network, based on the eigenvalues of symmetric part of local and coupling dynamics. Since this condition encompasses the Lyapunov spectrum, it is a sufficient condition on the stability \cite{Daleckii74}. Furthermore, we generalize the conventional setup where the linkage matrices are diagonal (often with binary components) to the case of an arbitrary linkage matrix, allowing multi-state and cross-state linkages, possibly with different strengths, which is now receiving more attention \cite{Martins01}\cite{Medvedev10}.

We then use the derived MSC to calculate a lower bound on the probability of stability for large Erd\"os-R\'enyi networks. We relate the condition of stability to dynamical characteristics of individual nodes and their coupling and structural properties of the Erd\"os-R\'enyi networks, namely, network size and randomness parameter.

\section{System Model}
Consider a network of $N$ identical nodes with identical coupling dynamics
\begin{equation}
\label{eq: SingMast}
\dot{\mathbf{x}}_i=\mathcal{F}(\mathbf{x}_i)-\sum_{\scriptscriptstyle j=1}^N{l_{ij}\mathcal{H}(\mathbf{x}_j)},
\end{equation}
where $\mathbf{x}_i\in\mathbb{R}^n,~ i=1,...,N,$ denotes the state vector of node $i$, and $\mathcal{F}(\mathbf{\cdot})$ and $\mathcal{H}(\mathbf{\cdot})$ denote the node and coupling dynamics, respectively. $\bL=[l_{ij}]$ is the Laplacian matrix of the network. Since  $\bL$ has zero row-sum, this network has a synchronization state, $\mathbf{x}_0$, which is the solution of local state equation, $\dot{\mathbf{x}}_0=\mathcal{F}(\mathbf{x_0})$, and $\mathbf{x}_{0}(0)=\mathbf{x}_{00}$. This equation also defines the synchronization manifold. To maintain the synchrony throughout the network, all $\mathbf{x}_i$ should converge to a synchronous state, $\mathbf{x}_0$. This means that all the modes traverse to the synchronization manifold should be damped out \cite{Pecora98}. Denote the deviation of the state of each node from the synchronous state by $\be_i=\mathbf{x}_i-\mathbf{x}_0$. If $\be_i$ are small, (\ref{eq: SingMast}) can be linearized around $\mathbf{x}_0$ as
\begin{equation}\label{eq: CellEq}
\dot{\be}_i=\bF\be_i-\sum_{\scriptscriptstyle j=1}^N l_{ij}\bH \be_j,
\end{equation}
where $\bF$ and $\bH$ are Jacobian matrices of $\mathcal{F}(\cdot)$ and $\mathcal{H}(\cdot)$ around $\mathbf{x}_0$, respectively. Note that in general, $\mathbf{x}_{0}$ is time dependent and so are the Jacobians. Here, for brevity of the presentation, we have dropped the explicit dependence of $\bF$ and $\bH$ on $\mathbf{x}_{0}$ (and therefore on time) from our notation. Combining (\ref{eq: CellEq}) for all \emph{i}, we obtain the unified dynamical equation of the whole network as
\begin{equation}
\dot{\be}=(\underbrace{\bI_N\otimes \bF-\bL\otimes \bH}_{\mbox{$\tilde{\bF}$}})\be,\label{eq: NetEq}
\end{equation}
where $\bI_N$ is the identity matrix of size $N$, $\otimes$ denotes the Kronecker product, and $\be=[\be_1^T~\be_2^T~\cdots\be_N^T]^T$ is the error vector with respect to $\mathbf{1}_{N}\otimes\mathbf{x}_{0}$, where $\mathbf{1}_{N}=[1~1~\cdots~1]_{N\times 1}$. Superscript $T$ denotes the transpose operator.

In the rest of the paper, we assume that the network is undirected, and therefore its Laplacian matrix is symmetric, i.e., $l_{ij}=l_{ji}$ for all $i,j$. 
\section{Network stability condition}
The network synchronization is exponentially stable around $\mathbf{x}_0$, if  $\tilde{\bF}$ is exponentially stable.  Due to zero row sum property of $\bL$, its smallest eigenvalue, $\mu_N$, is zero \cite{Mohar91}. The mode for $\mu_{N}=0$ in \eqref{eq: NetEq} represents the variation parallel to manifold and hence should be omitted in the study of stability for traversal exponents \cite{Pecora98}. Furthermore, multiplicity of zero in the set of network eigenvalues is one, if and only if the the graph is connected \cite{Mohar91}. The remaining modes in (\ref{eq: NetEq}) are traverse to the synchronization manifold and they decay exponentially if and only if the corresponding modes of the system governed by $\tilde{\bF}$ are exponentially stable.

Now, considering the system \eqref{eq: NetEq}, the norm of $\be$ can be bounded by (\cite{Leonov07} Thms. 1 and 2)
\begin{equation*}
\|\be(t)\|\leq \|\be(t_0)\|e^{\frac1 2\int_{t_0}^{t}\lambda_{1}\left(\tilde{\bF}(\tau)+\mathbf{\tilde{\bF}}^{T}(\tau)\right)+\epsilon ~d\tau},
\end{equation*}
for all $t\geq t_0$ and some $\epsilon>0$. Here $\lambda_1(.)$ denotes the largest eigenvalue of the argument. We note that if $\mathbf{\tilde{\bF}}+\mathbf{\tilde{\bF}}^T+\epsilon \bI\prec\b0$ for all $t\ge t_0$, the exponent approaches negative infinity as $t$ approaches infinity and the system is stable. Thus, a sufficient condition on stability is
\begin{equation}
\exists \epsilon>0\,\,\,\mbox{s.t.}\,\,\,\mathbf{\tilde{\bF}}(t)+\mathbf{\tilde{\bF}}^T(t)+\epsilon \bI\prec\b0,~\forall t\ge t_{0},\nonumber
 \end{equation}
here $\prec \b0$ denotes negative definiteness of a matrix. This condition guaranties damping of traverse modes and can be used as stability criteria in the study of dynamical systems \cite{Pecora98}.

Since $\tilde{\bF}$ can be written in block Jordan form, the equivalent sufficient condition on asymptotic stability for \eqref{eq: NetEq} yields
\begin{equation}\label{eq: MSC2}
\bF+\bF^T-\mu_i  \left(\bH+\bH^T\right)+\epsilon \bI\prec0,~\forall t\ge t_{0},
\end{equation}
for $1\le i\leq N-1$. Hence, $\tilde{\bF}$ is exponentially stable if and only if $\bF-\mu_i  \bH$ is stable for all $i=1,~2,~\cdots,~N-1$, where $\mu_{i}$ is the $i$th largest eigenvalue of $\bL$ (see appendix for propf).
The condition in (\ref{eq: MSC2}) requires the largest eigenvalue of symmetric part of $\bF-\mu_i  \bH$ to be strictly negative, and not approaching zero from below. Here we have used the assumption that \bL~is symmetric.
 Thus, $\mu_i$ are real and non-negative\footnote{In the case of directed network, $\mu_{i}\bH^{T}$ is replaced with $\mu_{i}^{*}\bH^{\dag}$, where superscripts $*$  and $\dag$ denote Hermitian and complex conjugate operators, respectively.}. In the case of complex $\bF \mbox{ and } \bH$, Hermitian operator replaces the transpose operator.

Now we can use Weyl's inequalities to further explore (\ref{eq: MSC2}). The Weyl inequalities provide upper and lower bounds on the eigenvalues of sum of Hermitian matrices. Assume that $\mathbf{B}$ and $\mathbf{C}$ are $n \times n$ Hermitian matrices, and $1\le j,k,j+k-n\le n$, then Weyl's inequalities state that (\cite{So94} Thm. 1.3)
\begin{eqnarray}
\lambda_{j+k-1}(\mathbf{B+C}) & \leq & \lambda_{j}(\mathbf{B})+\lambda_k(\mathbf{C}),\label{eq:honeycomb2}
\end{eqnarray}
and
\begin{eqnarray}
\lambda_{j+k-n}(\mathbf{B+C}) \ge \lambda_{j}(\mathbf{B})+\lambda_{k}(\mathbf{C}),\label{eq:honeycomb1}
\end{eqnarray}
where $\lambda_k(.)$ denotes the $k$th largest eigenvalue of the argument. Utilizing (\ref{eq:honeycomb2}) with $j+k-1=1$ and (\ref{eq:honeycomb1}) with $j+k-n=1$ yields
\begin{eqnarray}
\lambda_{1}(\mathbf{B+C}) & \leq & \lambda_{1}(\mathbf{B})+\lambda_{1}(\mathbf{C}),\label{eq:honeycomb21}
\end{eqnarray}
and
\begin{eqnarray}
\lambda_{1}(\mathbf{B+C}) \ge \lambda_{j}(\mathbf{B})+\lambda_{n-j+1}(\mathbf{C}),\label{eq:honeycomb11}
\end{eqnarray}
for $1\le j \le n$.
Now, since $\bF+\bF^T$ and $-\mu_i (\bH+\bH^T)$ are Hermitian, we can use (\ref{eq:honeycomb21}) and (\ref{eq:honeycomb11}) to bound $\lambda_{1}(\bF+\bF^T-\mu_i \bH-\mu_i\bH^T)+\epsilon$ from both sides:
\begin{align}
\lambda_{1}(\bF+\bF^T-\mu_i \bH-\mu_i\bH^T)+\epsilon\leq ~&\lambda_1(\bF+\bF^T)+\lambda_1(-\mu_i\bH-\mu_i\bH^T)+\epsilon\nonumber\\
&=\lambda_1(\bF+\bF^T)-\mu_i\lambda_n(\bH+\bH^T)+\epsilon,\label{eq: BoN1}
\end{align}
and
\begin{align}
\lambda_{1}(\bF+\bF^T-\mu_i \bH-\mu_i\bH^T)+\epsilon
\ge  ~&\lambda_{j}(\bF+\bF^T)+\lambda_{n-j+1}(-\mu_i\bH-\mu_i\bH^T)+\epsilon\nonumber\\
&=\lambda_{j}(\bF+\bF^T)-\mu_i\lambda_{j}(\bH+\bH^T)+\epsilon,\label{eq: BoN2}
\end{align}
for $1\le j \le n$.

To obtain a sufficient condition on stability we now force the upper bound in (\ref{eq: BoN1}), to be negative
\begin{equation}
\lambda_1(\bF+\bF^{T})- \min_{i}\{\mu_{i}\lambda_n(\bH+\bH^{T})\}+\epsilon<0.\label{eq: UB}
\end{equation}
for $i=1,~\cdots,~N-1$ and some $\epsilon>0$. Since $\lambda_n(\bH+\bH^{T})$ maybe positive or negative, the second term in (\ref{eq: UB}) reduces to $ \min\{\mu_{N-1}\lambda_n(\bH+\bH^{T}),\mu_{1}\lambda_n(\bH+\bH^{T})\}$. Thus, a sufficient condition on stability is
\begin{equation}\label{eq: stabcond1}
\begin{array}{ll}
\mu_{N-1}>\mu^{\dag} & \mbox{if}\,\,\,\lambda_n(\bH+\bH^{T})>0\\
\mu_{1}<\mu^{\dag} & \mbox{if}\,\,\,\lambda_n(\bH+\bH^{T})<0
\end{array}
\end{equation}
for some $\epsilon>0$, where $\mu^{\dag}={\lambda_1(\bF+\bF^{T})+\epsilon}/{\lambda_n(\bH+\bH^{T})}$.

We can also derive a necessary condition for \eqref{eq: MSC2} by forcing the lower bound in (\ref{eq: BoN2}) to be negative, or
\begin{equation}
0\geq\max_{i,j}\left\{\lambda_j(\bF+\bF^{T})-\mu_{i}\lambda_j(\bH+\bH^{T})\right\}+\epsilon.\label{eq: instab cond1}
\end{equation}
Let $k$ be the index of smallest positive eigenvalue of $\bH+\bH^T$. Then (\ref{eq: instab cond1}) reduces to
\begin{equation}\label{eq: instabcond1}
\begin{array}{ll}
\mu_{N-1}\geq\mu_{N-1}^{\ddag} \\
\mu_{1}\leq \mu_{1}^{\ddag},
\end{array}.
\end{equation}
where
\[\mu_{N-1}^{\ddag}\triangleq\max_{j\le k}\frac{\lambda_j\left(\bF+\bF^{T}\right)+\epsilon}{\lambda_j\left(\bH+\bH^{T}\right)},\]
and \[\mu_{1}^{\ddag}\triangleq\max_{j>k}\frac{\lambda_j\left(\bF+\bF^{T}\right)+\epsilon}{\lambda_j\left(\bH+\bH^{T}\right)}.\]

In the case that $\lambda_j(\bH+\bH^{T})=0$, if $ \lambda_j(\bF+\bF^{T})>0$ then the condition \eqref{eq: MSC2} is not satisfied, and if $ \lambda_j(\bF+\bF^{T})<0$, the corresponding condition can be eliminated.

We note that, (\ref{eq: instabcond1}) is a necessary condition on \eqref{eq: MSC2}, which itself is a sufficient condition on stability. Therefore (\ref{eq: instabcond1}) does not reveal anything about the stability of the network. However, since (\ref{eq: stabcond1}) and (\ref{eq: instabcond1}) sandwich \eqref{eq: MSC2}, (\ref{eq: instabcond1}) provides some information regarding how close \eqref{eq: MSC2} and (\ref{eq: stabcond1}) are.

Having developed the bounds on condition \eqref{eq: MSC2}, namely \eqref{eq: stabcond1} and \eqref{eq: instabcond1}, we now proceed to relate them to the degree properties of the network. To do this, we employ following inequalities known for symmetric Laplacian matrices \cite{Mohar91}
\begin{equation}
\mu_{N-1} \leq \frac{N}{N-1}d_{\min}, \nonumber
\end{equation}
and
\begin{equation}
\mu_{1} \leq \frac{N}{N-1}d_{\max},\nonumber
\end{equation}
where $d_{max}$ and $d_{min}$ denote the maximum and minimum node degrees, respectively. Using these, and (\ref{eq: stabcond1}), the stability condition can be also expressed as
\[\begin{array}{ll}d_{\min}\ge d_{\min}^{\dag}&\mbox{if}~\lambda_n(\bH+\bH^{T})>0\\d_{\max}\le d_{\max}^{\dag} & \mbox{if}~\lambda_n(\bH+\bH^{T})<0\end{array},\]
where
\begin{equation}
d_{\min}^{\dag}\triangleq\frac{N-1}{N}\mu_{N-1}^{\dag}\,\,\,\nonumber
\end{equation}
and\begin{equation}
d_{\max}^{\dag}\triangleq\frac{N-1}{N}\mu_{1}^{\dag} .\label{eq: stabconDegree}
\end{equation}
Similarly, necessary conditions for \eqref{eq: MSC2} become
\begin{eqnarray*}d_{\min}\ge d_{\min}^{\ddag}\\d_{\max}\le d_{\max}^{\dag} ,\end{eqnarray*}
where
\begin{equation}
	d_{\min}^{\ddag}\triangleq\frac{N-1}{N}\mu_{N-1}^{\ddag}\nonumber
\end{equation}
and,
\begin{equation}
	d_{\max}^{\ddag}\triangleq \frac{N-1}{N}\mu_{1}^{\ddag},\nonumber
\end{equation}
From (12) one can draw the conclusion that in the network with low algebraic connectivity,
$\mu_{N-1}$, synchronization is difficult to achieve. This behavior is caused by the fact that the coupling is not strong enough to push/pull the oscillators to synchronous state. And from \eqref{eq: stabconDegree} we can see the other case of non-synchronization behavior occurs when (some) nodes have too many connections (condition on $\mu_{1}$). This phenomenon is known as synchronization quenching, where the coupling is so strong that eliminates the self-drive of (some of) the oscillators and consequently, the network cannot achieve synchrony \cite{Osipov97}.

\section{Probability of stability for Erd\"os-R\'enyi networks}
In the following, we investigate probability of stability of Erd\"{o}s-R\'{e}nyi networks \cite{Bollobas01}. For large Erd\"{o}s-R\'{e}nyi networks with randomness parameter $p$, we can use (\ref{eq: stabcond1}) and (\ref{eq: instabcond1}) to calculate the lower and upper bounds on the probability of \eqref{eq: MSC2} being satisfied. We recall that \eqref{eq: stabcond1} provides a lower bound on the probability of stability, whereas \eqref{eq: instabcond1} describes the closeness of this lower bound and the probability of \eqref{eq: MSC2}.

Since eigenvalues of any large randomly generated symmetric matrix follows the Wigner's semi-circular distribution \cite{Wigner58}, we can approximate the distribution of eigenvalues of Laplacian for an Erd\"os-R\'enyi network. By definition $\bL=\mbox{diag}(\mathbf{d})-\bA,$ where \bA~is the adjacency matrix of the network, and $\mathbf{d}$ is the degree sequence of the nodes. Also in large Erd\"os-R\'enyi network, we can approximate $\mbox{diag}(\mathbf{d})$ by $Np\, \bI$, the average degree of the network \cite{Bollobas01}. Hence, $\bL\approx Np\,\bI-\bA$. Thus the eigenvalues of \bL~have (approximately) the following distribution \cite{Wigner58}
\begin{equation}
p_{\mu}(x)=\left\{\begin{array}{ll}
                    \frac{1}{\pi \sqrt{N p (1-p)}}\sqrt{1-X^2} & |X|< 1\\
                    0 & \mbox{elsewhere}
                  \end{array}\right.,\nonumber
\end{equation}
where
\begin{equation}
X = \frac{x-Np}{2\sqrt{Np(1-p)}}. \nonumber
\end{equation}
The order statistics $\mu_{N-1}$ and $\mu_{1}$ have densities
\begin{equation}
p_{\mu_{N-1}}(x)=(N-1)[1-P_{\mu}(x)]^{N-2}p_{\mu}(x),\nonumber
\end{equation}
and
\begin{equation}
p_{\mu_{1}}(x)=(N-1)[P_{\mu}(x)]^{N-2}p_{\mu}(x),\nonumber
\end{equation}
where
\begin{eqnarray}
P_{\mu}(x)&= &\int^x_{(N-1)p-2\sqrt{N p (1-p)}}p_{\mu}(y)dy\nonumber\\
&= &1-\frac{1}{\pi}\cos^{-1}X+\frac{1}{\pi}X\sqrt{1-X^2}.\nonumber
\end{eqnarray}
Now, we can evaluate the probability of occurrence of each conditions given in \eqref{eq: stabcond1} and \eqref{eq: instabcond1}. The lower bound on the probability of stability attained from Wigner's approximation for \eqref{eq: stabcond1} is
\begin{equation}
P_{\small \mbox{W,LB}}=\left\{\begin{array}{ll}
1-\left[1-P_\mu\left(\mu^{\dag}\right)\right]^{N-1}&
\lambda_n(\bH+\bH^T)>0\\
\left[P_\mu\left(\mu^{\dag}\right)\right]^{N-1}&
\lambda_n(\bH+\bH^T)<0\end{array}\right.\label{eq: Wsuf}
\end{equation}
Similarly, an upper bound on the probability of \eqref{eq: MSC2} being satisfied can be derived by applying Wigner's approximation to (\ref{eq: instabcond1})
\begin{eqnarray}
P_{\small\mbox{W,UB}}  =  \frac{1}{2}\left(1-[1-P_\mu(\mu_{N-1}^{\ddag})]^{N-1}+P_\mu^{N-1}(\mu_{1}^{\ddag}) -[P_\mu(\mu_{1}^{\ddag})-P_\mu(\mu_{N-1}^{\ddag})]^{N-1} \right).\label{eq: Wnec}
\end{eqnarray}

Since these probabilities have relatively sharp roll-offs as a function of $N$ (see numerical results), we can use randomness values, $p_{\small\mbox{L}}\mbox{ and } p_{\small\mbox{U}}$, which yield $P_{\small\mbox{W,LB}}(p_{\small\mbox{L}})=1/2$ and $P_{\small\mbox{W,UB}}(p_{\small\mbox{U}})=1/2$, to study the synchrony trends as network parameters change. From (\ref{eq: Wsuf}) and (\ref{eq: Wnec}), we have
\begin{equation}
	 p_{\small\mbox{L}}\approx\frac{1}{N+4}\left[\frac{\lambda_1(\bF+\bF^T)+\epsilon}{\lambda_n(\bH+\bH^T)}+4\right], \nonumber
\end{equation}
and
\begin{equation}
	 p_{\small\mbox{U}}\approx\frac{1}{N+4}\left[\max_{j>k}\frac{\lambda_j(\bF+\bF^T)+\epsilon}{\lambda_j(\bH+\bH^T)}\right]. \nonumber
\end{equation}
Thus, the value of randomness, \emph{p}, required to have stable synchronous state decreases as $O(1/N)$.

\section{Numerical Results}
For a numerical example, we consider a network of R\"{o}ssler oscillators (with parameters $a_1 = 0.165$, $a_2 = 0.2$, and $a_3 = 10$) \cite{Sorrentino08} coupled through all of their states. This set of parameters results in a coherent chaotic oscillation, since $a_1<0.21$ \cite{Osipov97}. Therefore, the Jacobian of the oscillator can be computed as
\begin{equation}
\bF=\left[\begin{array}{ccc}0 & -1 & -1\\1 & 0.165 & 0 \\x_{3} & 0 &x_{1} -10\end{array}\right] \nonumber
\end{equation}
We also assume
\begin{equation}
\bH=\frac{c}{\sqrt{6}+\sqrt{3}+\sqrt{2}}\left[\begin{array}{ccc}\sqrt{2} & \sqrt{2} & \sqrt{2} \\0 &\sqrt{3} & \sqrt{3}\\0 & 0 & \sqrt{6}\end{array}\right].\nonumber
\end{equation}
where $c=\mbox{trace}(\bH)$, is the coupling strength.

\begin{figure}
\begin{center}
\includegraphics[width=5.3in]{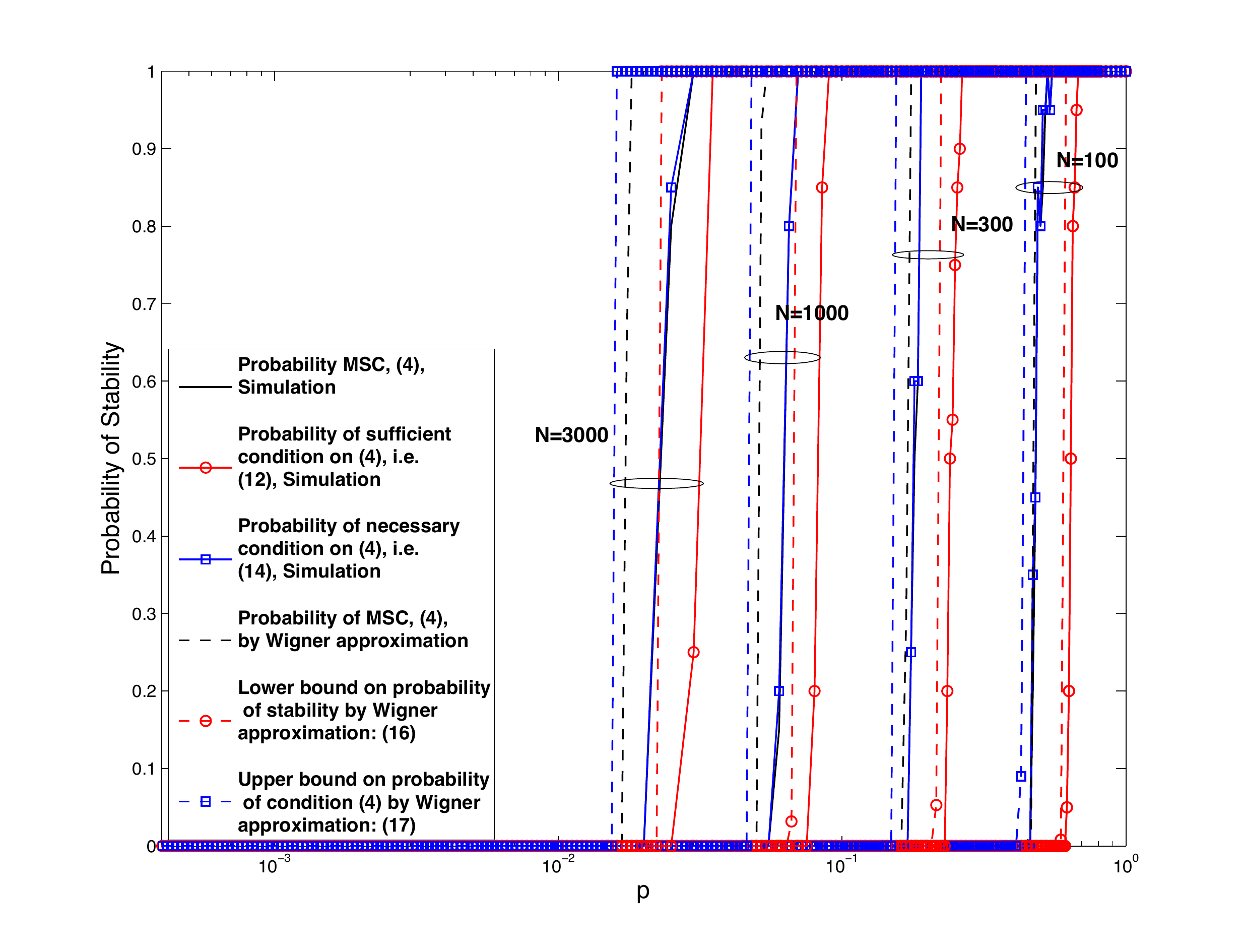}
\caption{Probability of stability as a function of $p$.}\label{fig: PNP}
\end{center}
\end{figure}

With this setting we calculate our results on the probability of stability of the network for different values of $N$, $p$, and $c$. To this end, we have considered $N$ trajectories starting from initial point $[s_{i}~ s_{j}~0]^{T}$ where $s_{j}$'s are selected uniformly from interval $[0.95,~1.05]$ and all the corresponding eigenvalues are calculated over average of $20$ cycles of initiated trajectories.

Fig. \ref{fig: PNP} shows the probability of the stability of the network as a function of network randomness, \emph{p}, for different network size, \emph{N}, and with coupling strength $c=1$. As it can be seen, probabilities of (\ref{eq: stabcond1}) and  (\ref{eq: instabcond1}) are close to that of (\ref{eq: MSC2}). Moreover, we observe that  the approximated probabilities provided by the Wigner's distribution of eigenvalues of the network are also reasonably close.

As Fig. \ref{fig: PNP} shows for positive definite $\bH+\bH^{T}$ in a large network if the average degree, $pN $, is above some \emph{threshold}, $ p_{L}N\approx\lambda_1(\bF+\bF^T)/\lambda_n(\bH+\bH^T) $, (in this example approximately $50$) the network becomes stable.
Note that in this particular numerical example, due to positive definiteness of $\bH+\bH^T$, only the transition from asynchrony to synchrony is observed.

\begin{figure}
\begin{center}
\includegraphics[width=5.3in]{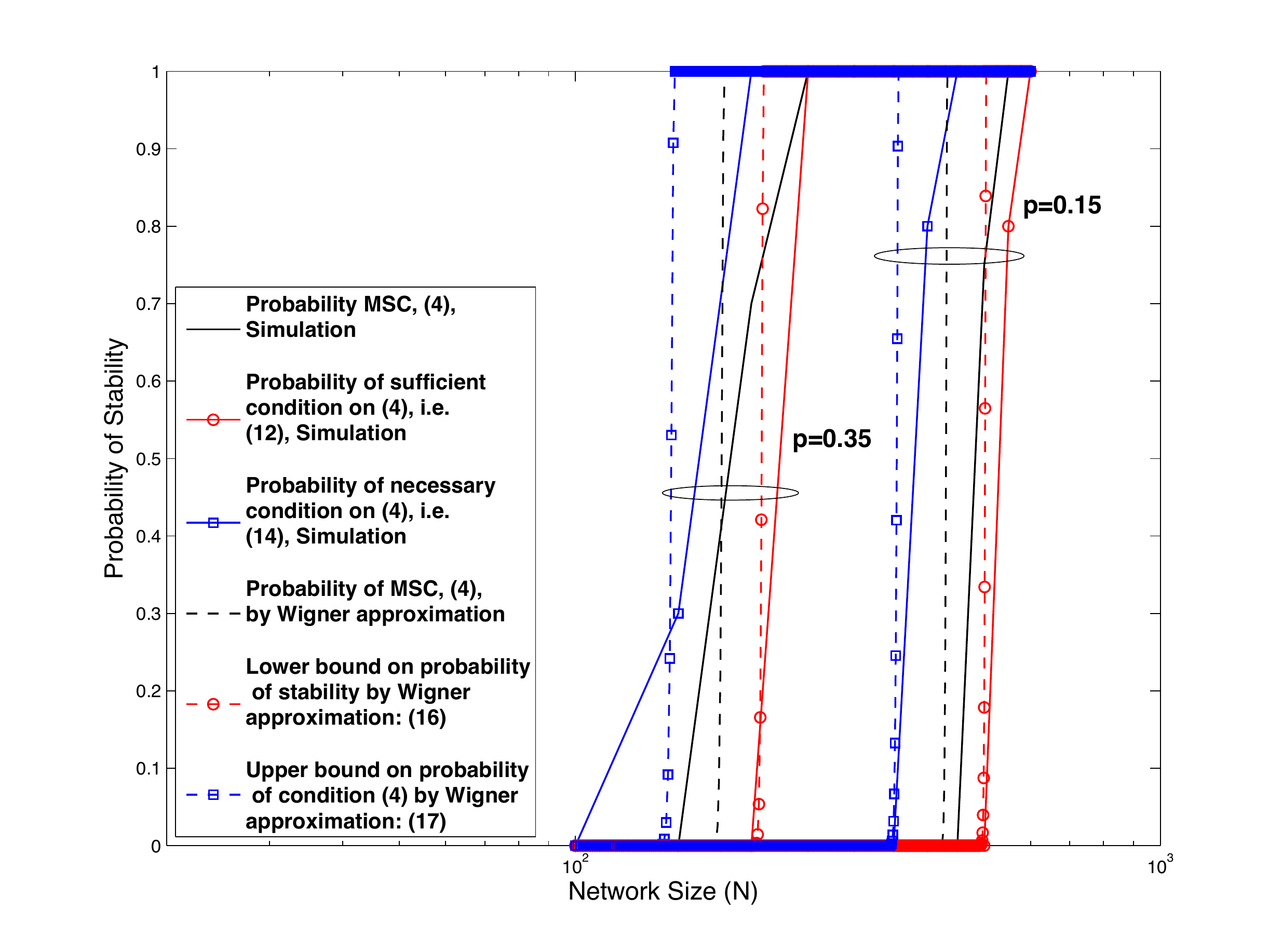}
\caption{Probability of stability as a function of $N$.}\label{fig: PPN}\end{center}
\end{figure}

The behavior of the network in the sense of its stability versus network size for several values of $p$ is shown in Fig. \ref{fig: PPN}. Once again we observe that the probability of stability suddenly increases as $pN$ crosses the threshold above.

\begin{figure}\begin{center}
\includegraphics[width=5.3in]{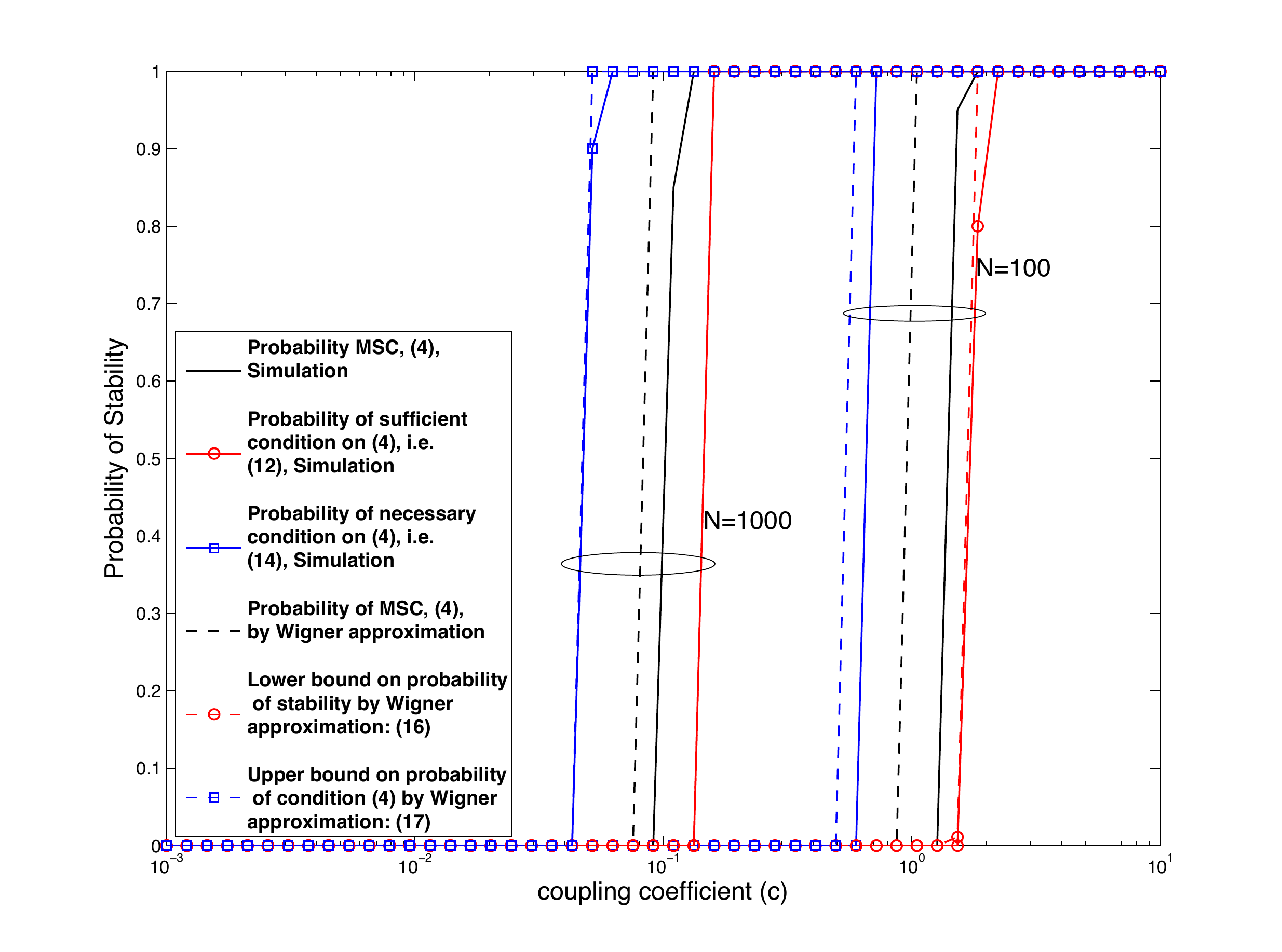}
\caption{Probability of stability as a function of $c$.}\label{fig: PNC}
\end{center}
\end{figure}

Fig. \ref{fig: PNC} shows that the stability of the network grows as the network size and coupling factor increase. Of course this is due to our choice of coupling dynamics which is positive definite and by increasing coupling strength, it provides stronger negative feedback to stabilize the network.

\section{Conclusion}
In conclusion, considering an alternative master stability condition, we have derived a sufficient condition of stability which is a function of the eigenvalues of network structure and symmetric parts of linearized local and coupling dynamics. Our condition relates the largest eigenvalues of the symmetric coupling and the symmetric local dynamics to stability conditions of the networks. For Erd\"os-R\'enyi network we have calculated a lower bound on the probability of stability. Then we have proceeded to calculated associated threshold value of randomness where the system starts to become stable as $p$ increases beyond this threshold. The reason for this phenomenon is that below the certain threshold, all or some of the nodes cannot achieve sufficient information exchange. As a result, those nodes cannot synchronize themselves with the rest of the network.

\appendix
\section{Appendix: stability of  $\dot{\be}=\tilde{\bF}\be$}\vspace{-0.2cm}
Since \bL is real and symmetric, it is unitarily and orthogonally diagonalizable (Ch. 5.4. [15]). That is, $\bL=\mathbf{Q}\boldsymbol{\Lambda}\mathbf{Q}^T$, where $\mathbf{Q}$ is unitary and $\boldsymbol{\Lambda}$ is diagonal \cite{Laub05}. Define $\mathbf{\bar{\be}}=(\mathbf{Q}^{T}\otimes \bI)\be$. Since $\be$ are mapped to $\mathbf{\bar{\be}}$ by a non-singular linear transformation ($\mathbf{Q}^{T}$ is unitary), their stabilities are equivalent. We have
\begin{equation}\label{eq: sys}
\dot{\bar{\be}}=(\mathbf{Q}^{T}\otimes \bI)\tilde{\bF}(\mathbf{Q}\otimes\bI)\mathbf{\bar{\be}},
\end{equation}
using the properties of the Kronecker product we have
\begin{eqnarray}
\dot{\bar{\be}}&=&(\mathbf{Q}^{T}\otimes \bI)(\bI\otimes\bF)(\mathbf{Q}\otimes\bI)\mathbf{\bar{\be}}+(\mathbf{Q}^{T}\otimes \bI)(\mathbf{QJQ}^{T}\otimes\bH)(\mathbf{Q}\otimes\bI)\mathbf{\bar{\be}}\nonumber\\
&=&(\mathbf{Q}^{T}\otimes \bF)(\mathbf{Q}\otimes\bI){\bar{\be}}+(\mathbf{JQ}^{T}\otimes\bH)(\mathbf{Q}\otimes\bI)\mathbf{\bar{\be}}\nonumber\\
&=&\left(\bI\otimes\bF+\boldsymbol{\Lambda}\otimes\bH\right){\bar{\be}}.\nonumber
\end{eqnarray}
The matrix $\bI\otimes\bF$ is block diagonal. Thus the resultant diagonal block matrix has $\bF-\mu_{i}\bH$ as its diagonal blocks. In other words, stability of $\tilde{\bF}$ is equivalent to stability of $\bF-\mu_i\bH $ for $i=1,\,\cdots,\,N$.

\end{document}